\documentclass[conference]{IEEEtran}

\usepackage{cite}
\usepackage{amsmath,amssymb,amsfonts}
\usepackage{algorithmic}
\usepackage{graphicx}
\usepackage{textcomp}
\usepackage{xcolor}
\usepackage{amsfonts}
\usepackage{url}
\usepackage{hyperref}
\usepackage{booktabs}
\usepackage{float}
\usepackage{amsmath}
\usepackage{array}

\def\BibTeX{{\rm B\kern-.05em{\sc i\kern-.025em b}\kern-.08em
    T\kern-.1667em\lower.7ex\hbox{E}\kern-.125emX}}
\begin{document}

\title{MalCodeAI: Autonomous Vulnerability Detection and Remediation via Language Agnostic Code Reasoning}

\author{\IEEEauthorblockN{1\textsuperscript{st} Jugal Gajjar}
\IEEEauthorblockA{\textit{Computer Science Department} \\
\textit{The George Washington University}\\
Washington D.C, USA \\
jugal.gajjar@gwu.edu}
\and
\IEEEauthorblockN{2\textsuperscript{nd} Kamalasankari Subramaniakuppusamy}
\IEEEauthorblockA{\textit{Computer Science Department} \\
\textit{The George Washington University}\\
Washington D.C, USA \\
kamalasankaris@gwu.edu}
\and
\IEEEauthorblockN{3\textsuperscript{rd} Noha El Kachach}
\IEEEauthorblockA{\textit{Computer Science Department} \\
\textit{The George Washington University}\\
Washington D.C, USA \\
noha.elkachach@gwu.edu}
}

\maketitle

\begin{abstract}
The growing complexity of cyber threats and the limitations of traditional vulnerability detection tools necessitate novel approaches for securing software systems. We introduce MalCodeAI, a language-agnostic, multi-stage AI pipeline for autonomous code security analysis and remediation. MalCodeAI combines code decomposition and semantic reasoning using fine-tuned Qwen2.5-Coder-3B-Instruct models, optimized through Low-Rank Adaptation (LoRA) within the MLX framework, and delivers scalable, accurate results across 14 programming languages. In Phase 1, the model achieved a validation loss as low as 0.397 for functional decomposition and summarization of code segments after 200 iterations, 6 trainable layers, and a learning rate of $2 \times 10^{-5}$. In Phase 2, for vulnerability detection and remediation, it achieved a best validation loss of 0.199 using the same number of iterations and trainable layers but with an increased learning rate of $4 \times 10^{-5}$, effectively identifying security flaws and suggesting actionable fixes. MalCodeAI supports red-hat-style exploit tracing, CVSS-based risk scoring, and zero-shot generalization to detect complex, zero-day vulnerabilities. In a qualitative evaluation involving 15 developers, the system received high scores in usefulness (mean 8.06/10), interpretability (mean 7.40/10), and readability of outputs (mean 7.53/10), confirming its practical value in real-world development workflows. This work marks a significant advancement toward intelligent, explainable, and developer-centric software security solutions.
\end{abstract}


\section{Introduction}
\label{intro}
As software continues to permeate every layer of modern society—from critical infrastructure and healthcare systems to finance and consumer applications—ensuring its security has become a global imperative. Despite the rapid growth of secure development practices, software vulnerabilities remain pervasive. According to the National Vulnerability Database (NVD), thousands of new vulnerabilities are reported each year, many of which are exploited by attackers before they can be patched, a stark indication that traditional tools for vulnerability detection are struggling to keep pace with the evolving threat landscape \cite{nvd-dashboard}.

Conventional methods such as static analysis, dynamic analysis, and symbolic execution have been foundational in identifying software vulnerabilities. However, these techniques are inherently limited. Static analyzers often produce high false-positive rates and lack contextual understanding of a program’s behavior \cite{harzevili2023}, while dynamic analysis is time-consuming and constrained by execution paths and test inputs \cite{alashjaee2019}. Furthermore, many modern malware and exploits are designed to evade these traditional techniques using obfuscation, polymorphism, and context-aware logic.

The advent of transformer-based models and large language models (LLMs), such as CodeBERT \cite{codebert}, GraphCodeBERT \cite{graphcodebert}, and GPT-based variants \cite{chen2021}, has introduced a paradigm shift in how source code can be analyzed. These models learn from vast corpora of code to capture both syntactic patterns and semantic meaning, enabling sophisticated tasks like code summarization, generation, and vulnerability detection. However, these systems are often constrained to known vulnerabilities, specific programming languages, or narrow datasets like Juliet \cite{juliet} or Devign \cite{devign}, limiting their generalizability to real-world, zero-day scenarios.

To address these limitations, we propose MalCodeAI, a language-agnostic, multi-stage AI pipeline for detecting and remediating vulnerabilities in source code. The system is composed of two primary models: (1) a componentization model that segments large codebases into logically independent functional units, and (2) a vulnerability detection and remediation model that performs semantic reasoning to identify threats, simulate potential exploits, and recommend security fixes. The models are fine-tuned using LoRA (Low-Rank Adaptation) \cite{lora} on top of Qwen2.5-Coder-3B-Instruct \cite{qwen25} to ensure scalability, cross-language support, and high contextual accuracy. Other contemporaneous models such as StarCoder 2 \cite{starcoder2}, DeepSeek-Coder V2 \cite{deepseekcoder}, and Llama 3 \cite{llama3} also show promising results in code intelligence tasks; however, Qwen2.5 was selected for its balance of openness, efficiency, and code reasoning performance.

Unlike prior work, MalCodeAI integrates red-hat-style exploit simulation, explainability, and zero-shot generalization across 14 popular programming languages. It bridges the gap between academic vulnerability detection research and real-world developer needs by offering detailed CVE-like scoring, behavioral annotations, and secure coding recommendations—transforming it from a diagnostic tool into an autonomous security assistant.

By enabling both proactive and reactive detection of vulnerabilities, MalCodeAI not only mitigates risk but also contributes toward the broader vision of secure-by-design software development. This paper explores the architecture, training, evaluation, and impact of this system, alongside its potential role in advancing intelligent cybersecurity solutions.

\section{Related Work}
\label{rel_work}
Traditional approaches to software anomaly and vulnerability detection largely relied on signature-based methods, which detect known threats by matching code patterns against a database of malicious signatures \cite{techslang}. These methods, widely used in antivirus systems, were effective for familiar threats but failed to generalize to zero-day exploits, AI-driven malware, and Advanced Persistent Threats (APTs) \cite{kothamali2022}. This led to a growing need for more advanced, adaptive security systems.

To address these challenges, static analysis techniques emerged, aiming to identify security vulnerabilities during the early stages of software development. Tools like SonarQube \cite{sonarqube} and Checkmarx \cite{checkmarx} perform rule-based scanning over Abstract Syntax Trees (ASTs) to catch issues such as buffer overflows and SQL injection. However, these tools often produce high false-positive rates and struggle with context-sensitive vulnerabilities \cite{harzevili2023}. More recently, graph-based approaches like Code Property Graphs (CPGs) and Control Flow Graphs (CFGs) have been used to structurally model program behaviors, improving detection accuracy. Frameworks such as QVoG compress CPGs to reduce memory usage and use parallelism to enhance performance, outperforming traditional tools like Joern and CodeQL on large-scale codebases \cite{liu2024}. However, they remain limited in language support.

In contrast, dynamic analysis techniques examine applications at runtime, enabling the discovery of vulnerabilities such as memory leaks and race conditions. Tools like Burp Suite \cite{burpsuite} and OWASP ZAP \cite{zap} exemplify this category. While dynamic analysis offers practical insights by observing application behavior, it requires complex execution environments and still suffers from incomplete code path coverage \cite{alashjaee2019}.

To overcome these limitations, researchers have explored machine learning (ML) and deep learning (DL) for vulnerability detection. Early ML-based approaches, such as those built on the EMBER dataset \cite{ember}, improved detection of known threats but lacked robustness against adversarial or zero-day samples \cite{kolosnjaji2018}. DL-based systems like VulDeePecker \cite{vuldeepecker} used BiLSTMs over code gadgets to identify vulnerabilities, demonstrating superior accuracy but demanding large labeled datasets and high computational resources \cite{le2020}. Models like GCN \cite{gcn}, which applied Convolutional Graph Neural Networks to ASTs for defect classification in Java, offered binary classification with improved performance but limited generalization \cite{gcn2022}. To address these issues, BGNN4VD combined ASTs, CFGs, and Data Flow Graphs (DFGs) with a bidirectional GNN and CNN architecture, achieving higher precision and F1 scores on C/C++ datasets \cite{bgnn4vd}.

The emergence of transformer-based models revolutionized NLP and subsequently code understanding. Pioneered by Vaswani et al. \cite{attention}, the transformer architecture laid the foundation for pre-trained models like BERT \cite{bert}, CodeBERT \cite{codebert_findings}, and CodeT5 \cite{codet5}, which have demonstrated impressive performance in tasks like code summarization, bug detection, and code completion. In the realm of security, models such as MalBERT \cite{malbert} and ViT4Mal \cite{vit4mal} applied transformer architectures to malware detection, leveraging byte-level and API call patterns. Other notable contributions include VulBERTa, which was trained on C/C++ code and showed competitive results \cite{vulberta}. However, despite these advances, transformer-based models often rely heavily on high-quality, curated datasets, are vulnerable to adversarial perturbations, and lack effective mechanisms to generalize to unseen, zero-day threats \cite{transformersurvey}.

Unlike prior works, MalCodeAI introduces a novel two-stage pipeline that first decomposes code into functional components and then applies semantic vulnerability detection, red-hat-style exploit reasoning, and automatic remediation. It expands on prior efforts by supporting 14 programming languages, enabling component-level explainability, and simulating zero-shot generalization through fine-tuned instruction-based LLMs. Additionally, the use of LoRA fine-tuning ensures adaptability with fewer computational resources, a step forward from monolithic pretraining approaches.

By merging program analysis, natural language understanding, and reasoning capabilities, MalCodeAI addresses the gaps in scalability, generalizability, and practical usability found in prior research.


\section{Methodology}
\label{method}

We present MalCodeAI, a novel framework for the automated detection and remediation of malicious code segments, leveraging large language models (LLMs) and a multi-stage pipeline. This methodology integrates advanced semantic analysis, security risk scoring, exploit reasoning, and explainability mechanisms. The framework is designed to be highly adaptable, supporting 14 programming languages, and is capable of zero-shot generalization to previously unseen codebases.

\subsection{Code Decomposition and Semantic Understanding}

The initial step in the pipeline involves decomposing the input code into independent functional components, facilitating detailed analysis, and ensuring that each piece of code can be assessed for potential vulnerabilities. This decomposition is achieved by a fine-tuned version of the Qwen2.5-Coder-3B-Instruct model \cite{qwen25}, which is trained to segment code into logical units while generating natural language descriptions of their purpose. The segmentation process is guided by syntax-aware token-level cues, allowing the system to preserve semantic boundaries. The fine-tuning process utilizes the MLX framework \cite{mlx} with LoRA (Low-Rank Adaptation) \cite{lora}, optimizing the model for minimal computational overhead without compromising accuracy. This component segmentation allows for a granular examination of the code, enabling the detection of abnormal or malicious behaviors at a component level.

\subsection{Preliminary Risk Assessment and CVSS Scoring}

Each decomposed code component is assigned an initial Common Vulnerability Scoring System (CVSS) score \cite{cvss} to prioritize components for further analysis. This score is calculated based on static code patterns, API usage, control flow deviations, and other heuristics indicative of common vulnerabilities. Components that exhibit high-risk patterns are flagged for deeper inspection, while lower-risk components are deprioritized, enabling the system to scale efficiently and focus computational resources on potentially dangerous code.

\subsection{Deep Security Analysis and Exploit Reasoning}

Following the preliminary risk assessment, the identified high-risk code components undergo a second round of analysis, utilizing the same fine-tuned Qwen2.5 model \cite{qwen25}. In this stage, the model applies advanced reasoning techniques to evaluate whether the components contain exploitable vulnerabilities such as backdoors, logic bombs, or privilege escalation points. The model simulates potential attack paths to assess how adversaries might exploit these vulnerabilities. MalCodeAI is fine-tuned on a custom dataset comprising labeled malicious and benign code snippets, focused on detecting ten critical vulnerability types (e.g., SQL Injection, Cross-Site Scripting, Remote Code Execution). This analysis provides an understanding of the threat landscape, going beyond basic pattern matching to capture complex attack vectors.

\subsection{Explainability and Red-Hat Style Exploit Tracing}

A key feature of MalCodeAI is its emphasis on explainability. For each vulnerability detected, the system generates a red-hat style exploit trace, which tells how an attacker could potentially exploit the identified vulnerability. This trace includes step-by-step descriptions of the attack path, including entry points, escalation techniques, and the ultimate impact of the exploitation. This enhances transparency and fosters trust among developers, making the model’s decisions more interpretable and actionable. The model’s reasoning is grounded in semantic understanding enabled by its pretraining via the Fill-in-the-Middle (FIM) technique on trillions of code tokens \cite{qwen25}.

\subsection{Automated Remediation Generation}

To assist developers in mitigating vulnerabilities, MalCodeAI generates automated remediation suggestions for each identified issue. These suggestions are formulated based on secure coding practices and are tailored to the specific vulnerabilities detected. The model leverages its instruction-following capabilities, fine-tuned through examples during training, to recommend changes that fix the issue while preserving the original functionality of the code. This step is critical for transforming the detection process into actionable and practical security interventions.

\subsection{Training and Optimization}

The fine-tuning of the Qwen2.5-Coder-3B-Instruct model \cite{qwen25} was performed using a carefully curated dataset containing labeled malicious and benign code snippets across 14 programming languages, including but not limited to Python, Java, C, C++, Rust, Go, and Scala. The LoRA fine-tuning configuration employed six trainable layers, a learning rate of $2 \times 10^{-5}$, and a maximum token sequence length of 3072 tokens. The optimization focused on improving detection accuracy while ensuring that the model operates efficiently in terms of both computational resources and memory usage.

\subsection{Multi-Language and Zero-Shot Generalization}

One of the distinguishing features of MalCodeAI is its multi-language support. The framework is designed to handle a wide range of programming languages, enabling it to detect vulnerabilities in both widely used and niche programming environments. Additionally, MalCodeAI is equipped with a zero-shot generalization capability, which allows the model to analyze code written in new, unseen languages or frameworks. This is achieved by leveraging the pre-trained model’s ability to generalize from prior knowledge and adapt to novel code structures, ensuring its applicability across diverse use cases, including proprietary or domain-specific languages.

\begin{figure*}[!t]
\centering
\includegraphics[width=0.8\textwidth, height=2in]{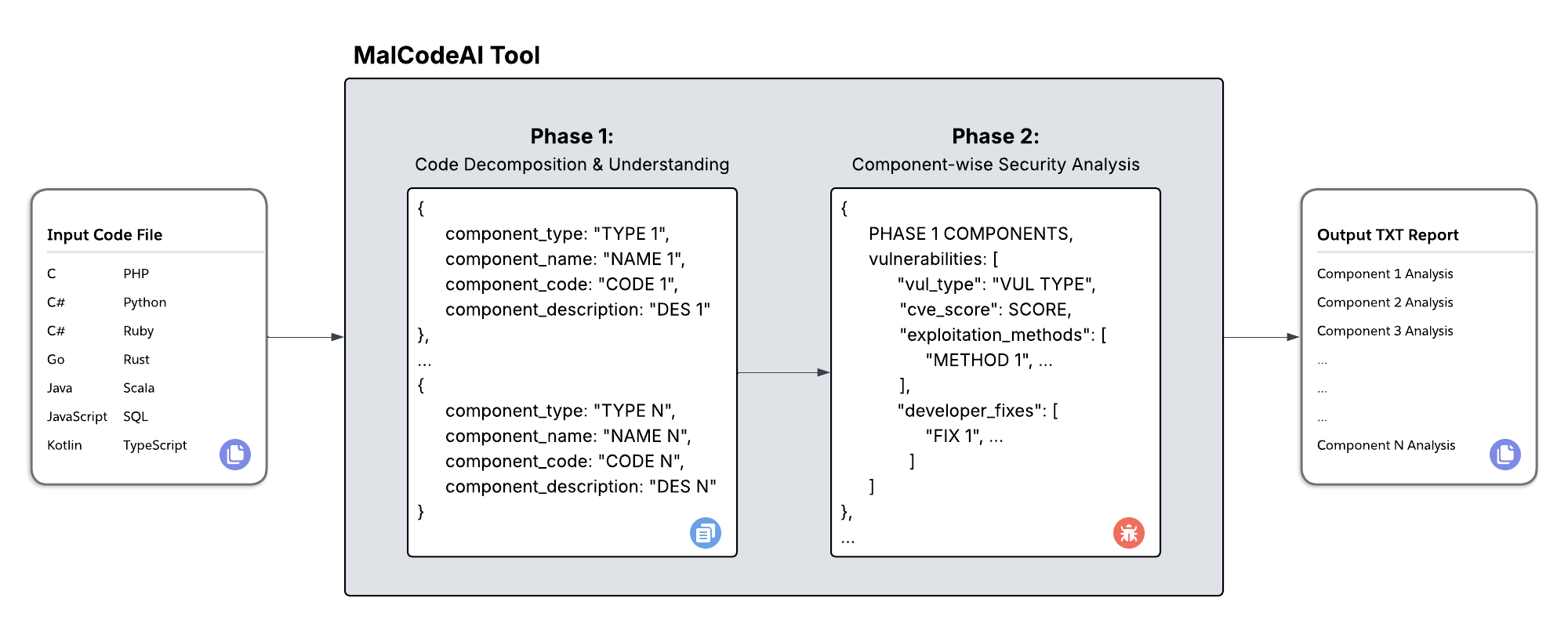}
\caption{Visual representation of the MalCodeAI pipeline. The tool processes a source code file through two primary phases. In Phase 1, the code is decomposed into functional components such as functions and classes, and each is enriched with semantic descriptions. In Phase 2, each component undergoes deep security inspection to identify vulnerabilities, simulate potential exploit paths, and suggest secure remediation strategies. The output is a component-wise report suitable for developer consumption and security audit documentation.}
\label{fig:malcodeai_architecture}
\end{figure*}

To further clarify the structure and data flow of MalCodeAI, Figure~\ref{fig:malcodeai_architecture} illustrates the complete pipeline from input code file to final output report. This diagram encapsulates the sequential processing steps described above, highlighting the two-phase architecture and how each code component is analyzed in isolation to support modular, explainable security insights.

\section{Experiments}
\label{experiments}
To evaluate the performance of the MalCodeAI system, we conducted a series of experiments focused on fine-tuning the Qwen2.5-Coder-3B-Instruct model \cite{qwen25} using the LoRA \cite{lora} technique within the MLX framework \cite{mlx}. All training was performed on an Apple Silicon M4 Pro device equipped with 48GB unified memory, leveraging the optimized lightweight capabilities of the MLX framework. Each fine-tuning phase (Phase 1 and Phase 2) required approximately 3 to 4 hours of execution time, utilizing a peak memory of 37GB. The experiments were conducted in two main phases, corresponding to the two critical components of the pipeline: code decomposition and vulnerability detection.

\subsection{Phase 1: Code Decomposition and Summarization}

In the first phase, the model was fine-tuned to break down source code files into independent components and generate meaningful descriptions. To optimize performance, we experimented with variations in key hyperparameters: number of iterations, trainable layers, and learning rate.


We began by varying the number of training iterations. Initially set to 100, this was increased to 200 to observe its impact. Results showed clear improvement, with validation loss dropping from 0.865 to 0.399.


Next, we tested different configurations for the number of trainable layers. The model performed best with 6 trainable layers; reducing them increased validation loss—0.470 at 6 layers, 0.504 at 5, and 0.555 at 4—demonstrating that deeper models were more effective.

Finally, we adjusted the learning rate to assess its impact on model performance. The best results were observed with a learning rate of $2 \times 10^{-5}$, which produced the lowest validation loss of 0.397. Higher learning rates resulted in slightly less effective performance, indicating that the lower learning rate allowed for better convergence during training.

\begin{figure}[htbp]
\centerline{\includegraphics[width=\linewidth, height=2.2in]{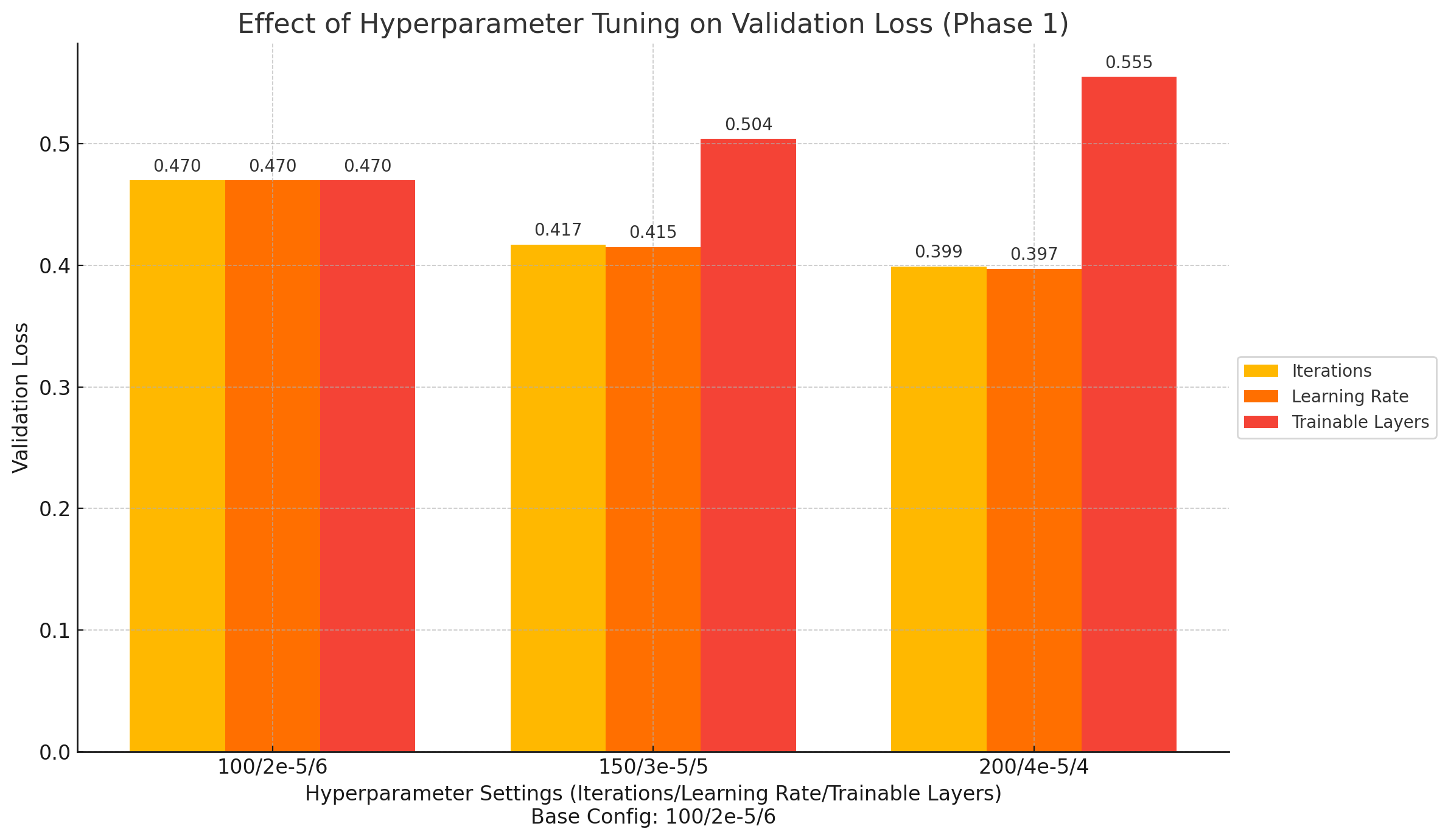}}
\caption{Grouped bar chart illustrating the impact of varying iterations, learning rate, and number of trainable layers on validation loss in Phase 1 experiments. The base model was fine-tuned using LoRA in MLX with default settings: 100 iterations, 6 trainable layers, learning rate of 2e-5, and 3072 maximum tokens. Increasing iterations and learning rate led to improved validation loss, while reducing the number of trainable layers degraded performance.}
\label{fig:phase1}
\end{figure}

Figure~\ref{fig:phase1} summarizes these findings, showing how each hyperparameter setting affects validation loss. The improvements in loss with more iterations and higher learning rates are clear, while a reduction in trainable layers consistently degrades performance.

After these adjustments, the final configuration for Phase 1 consisted of 200 iterations, 6 trainable layers, a learning rate of $2 \times 10^{-5}$, and a maximum token length of 3072. 

\subsection{Phase 2: Vulnerability Detection and Remediation}

The second phase aimed to fine-tune the model for vulnerability detection and remediation. Similar to Phase 1, we experimented with key hyperparameters: iterations, trainable layers, and learning rate.


The initial model showed a validation loss of 1.639 at 100 iterations. Increasing iterations improved performance, dropping the loss to 0.222 at 150 iterations and 0.203 at 200, indicating that extended training enhanced detection accuracy.


We also varied the number of trainable layers. As in Phase 1, the model performed best with 6 layers, and reducing the number of layers led to increase in validation loss from 0.325 at 6 layers to 0.442 at 5 layers and 0.585 at 4 layers.

Finally, we tested different learning rates. The learning rate of $4 \times 10^{-5}$ yielded the best results, with the lowest validation loss of 0.199. Higher or lower learning rates produced less effective models, reinforcing the importance of selecting the optimal learning rate for training.

\begin{figure}[htbp]
\centerline{\includegraphics[width=\linewidth, height=2.2in]{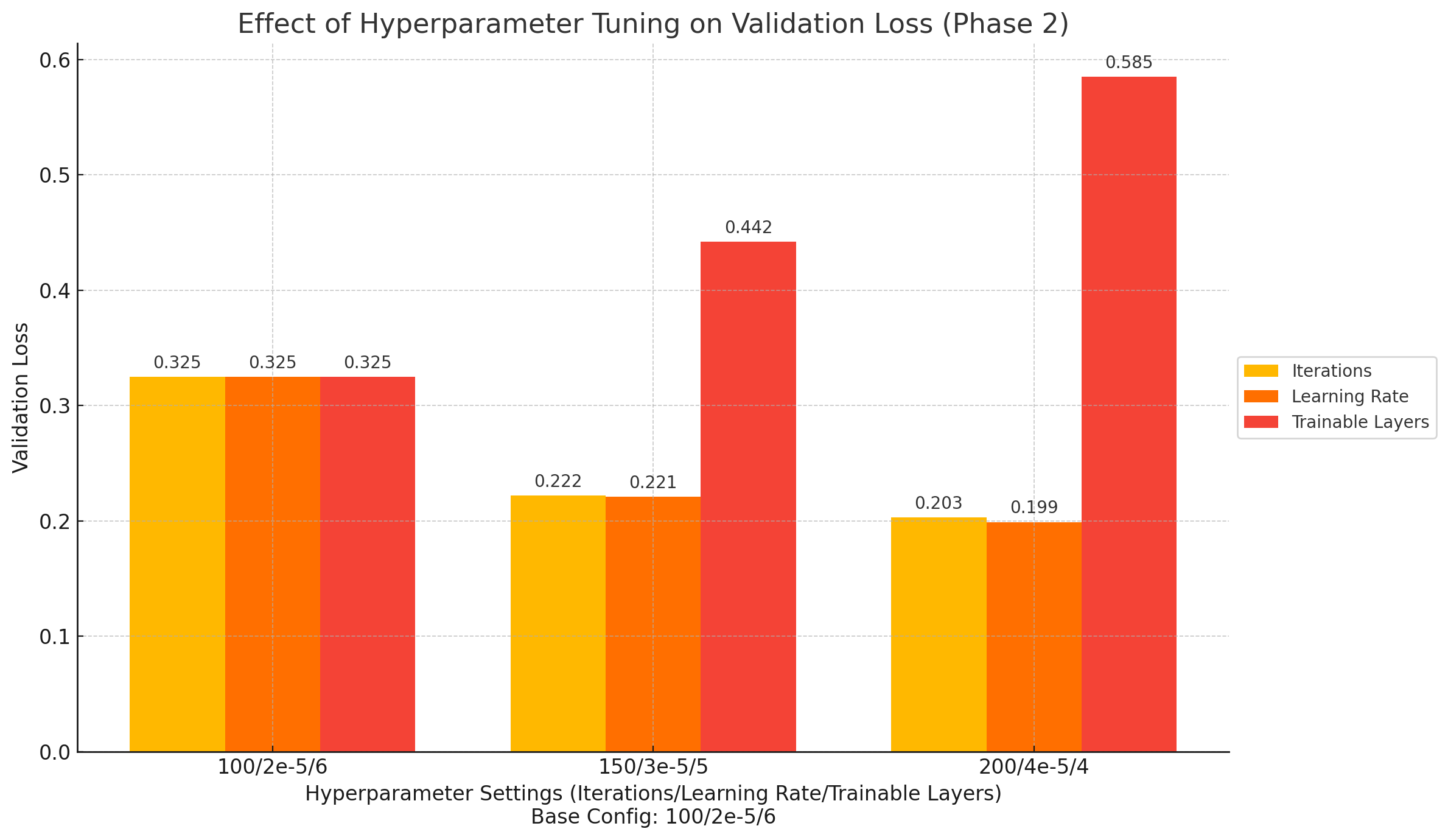}}
\caption{Grouped bar chart showing validation loss trends in Phase 2 experiments as each hyperparameter is varied while holding others constant. Similar to Phase 1, higher iteration counts and learning rates continue to yield better validation outcomes. However, reducing trainable layers again negatively impacts performance, reaffirming the importance of sufficient model capacity during fine-tuning.}
\label{fig:phase2}
\end{figure}

Figure~\ref{fig:phase2} illustrates these trends, confirming that increased iterations and optimal learning rates improve performance, while fewer trainable layers degrade it.

The final configuration for Phase 2 consisted of 100 iterations, 6 trainable layers, a learning rate of $4 \times 10^{-5}$, and a maximum token length of 3072. 

\section{Results}
\label{results}
We evaluated MalCodeAI through both quantitative and qualitative analyses to assess its performance in detecting and explaining code vulnerabilities across diverse programming languages. Moreover, we manually evaluated the remediation quality.

\subsection{Quantitative Results}

Phase 1 aimed to segment source code into independent components with meaningful summaries. Fine-tuning experiments showed that increasing training iterations and trainable layers improved performance. At 200 iterations, the model reached its lowest validation loss of 0.399. Among layer depths, six trainable layers yielded the best result (loss = 0.470), while five and four layers increased loss to 0.504 and 0.555, respectively. Learning rate experiments showed a minimum loss of 0.397 at $4 \times 10^{-5}$. The final Phase 1 configuration—200 iterations, six trainable layers, a learning rate of $2 \times 10^{-5}$, and a 3072-token limit—was selected for its stable convergence.

\begin{table}[htbp]
\caption{Phase 1 – Code Decomposition Results}
\begin{center}
\begin{tabular}{|l|p{1.8cm}|l|l|}
\hline
\textbf{Parameter} & \textbf{Tested Values} & \textbf{Best Value} & \textbf{Best Val. Loss} \\
\hline
Training Iterations & 100, 150, 200 & 200 & 0.399 \\
\hline
Trainable Layers & 4, 5, 6 & 6 & 0.470 \\
\hline
Learning Rate & Various (e.g., $2 \times 10^{-5}$) & $4 \times 10^{-5}$ & 0.397 \\
\hline
Max Token Length & -- & 3072 & -- \\
\hline
\end{tabular}
\label{tab:phase1}
\end{center}
\end{table}


Phase 2 addressed vulnerability detection and remediation. Starting from a base validation loss of 1.639, fine-tuning with 100 iterations, six trainable layers, and a $4 \times 10^{-5}$ learning rate reduced loss to 0.199. Fewer layers degraded performance, while higher learning rates improved convergence.

\begin{table}[htbp]
\caption{Phase 2 – Vulnerability Detection Results}
\begin{center}
\begin{tabular}{|l|p{1.8cm}|l|l|}
\hline
\textbf{Parameter} & \textbf{Tested Values} & \textbf{Best Value} & \textbf{Best Val. Loss} \\
\hline
Training Iterations & 100, 150, 200 & 200 & 0.203 \\
\hline
Trainable Layers & 4, 5, 6 & 6 & 0.325 \\
\hline
Learning Rate & Various (e.g., $2 \times 10^{-5}$) & $4 \times 10^{-5}$ & 0.199 \\
\hline
Max Token Length & -- & 3072 & -- \\
\hline
\end{tabular}
\label{tab:phase2}
\end{center}
\end{table}

\subsection{Qualitative Results}
To complement the quantitative findings, we collected structured feedback from 15 developers (a mixture of practitioners and CS students). They evaluated the system on three dimensions: usefulness, readability, and interpretability.

\begin{table}[htbp]
\caption{Developer Feedback Scores on Usefulness, Readability, and Interpretability ($n=15$)}
\begin{center}
\begin{tabular}{|l|c|c|c|l|}
\hline
\textbf{Dimension} & \textbf{Mean} & \textbf{Median} & \textbf{Mode} & \textbf{Observations} \\
\hline
Usefulness & 8.06 & 9.0 & 9.0 & Positive reception \\
\hline
Readability & 7.00 & 8.0 & 10 & Mixed views \\
\hline
Interpretability & 7.40 & 8.0 & 8.0 & Semi-clear descriptions \\
\hline
\end{tabular}
\label{tab:qualitative}
\end{center}
\end{table}

\begin{figure}[htbp]
\centerline{\includegraphics[width=\linewidth, height=1.35in]{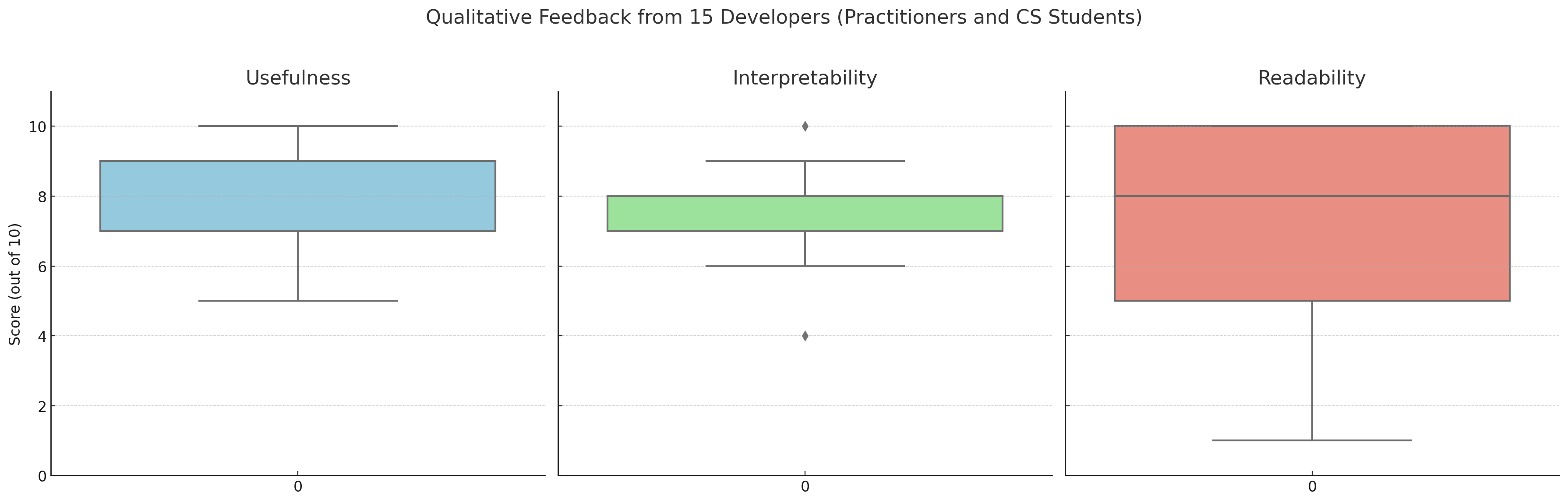}}
\caption{Box-and-swarm plot visualization of developer ratings on usefulness, readability, and interpretability ($n=15$). Usefulness and interpretability scores show high consistency and central tendency, while readability scores exhibit higher variance, suggesting varied stylistic preferences.}
\label{fig:qualitative-plot}
\end{figure}

Figure~\ref{fig:qualitative-plot} provides a complementary visualization of these ratings, capturing both the central tendencies and individual variances. While usefulness and interpretability received consistently high scores, readability showed a wider range, reflecting stylistic preferences and the balance between detail and conciseness.

\subsection{Remediation Quality}
To evaluate the quality of MalCodeAI’s remediation suggestions, we tested it on 20 manually crafted code files spanning four mini-projects in Python, C, Java, and JavaScript. These projects included one secure, one insecure, and two mixed-security setups. 

In the insecure project, 11 vulnerabilities were injected, of which 8 were correctly detected. Among these, 5 remediations were specific and actionable, while 3 were generic. For the mixed-security projects, 5 of 6 vulnerabilities were identified, with 4 specific and 1 generic fix. Notably, the system also flagged an overlooked vulnerability, later confirmed upon manual inspection. 

In the secure project, one false positive was flagged due to lack of runtime context, highlighting the limitations of static LLM-based evaluation. Overall, the model showed high recall and generated mostly relevant remediation suggestions, though some remained generic and constrained by static analysis.

These findings demonstrate that MalCodeAI is not only technically effective in detecting malicious code but also valuable as an assistive tool for developers in security-critical environments. Its explainability enhances developer trust, while its language-agnostic design ensures adaptability across ecosystems.

\section{Discussion \& Future Work}
\label{dis_and_fut_work}

While MalCodeAI shows strong potential as a language-agnostic, multi-stage vulnerability detection and remediation system, few limitations exist. First, its reliance on fine-tuned LLMs makes it sensitive to input formatting, length, and structure. Second, remediation suggestions are provided as text, which may not always translate to syntactically or semantically correct code. Lastly, MalCodeAI follows a static analysis approach and does not simulate exploit execution or dynamic behavior, potentially missing runtime vulnerabilities.


Future enhancements to MalCodeAI include integrating an agent-based sandbox for exploit simulation to uncover runtime vulnerabilities dynamically. Another direction is automated patch generation, enabling the system to both suggest and apply fixes. These additions would create a fully autonomous detection and remediation pipeline, enhancing real-world applicability and reducing manual effort for security teams.

\section{Conclusion}
\label{concl}
This project presents a transformer-based AI tool that effectively combines code parsing with vulnerability detection using fine-tuned large language models. By leveraging LLM reasoning, the system not only identifies potential vulnerabilities but also highlights plausible real-world exploit paths, providing developers with a deeper understanding of how their code could be exploited. In addition to detection, the tool delivers actionable remediation strategies, equipping developers with practical solutions to address the issues. By integrating security analysis early in the software development life cycle (SDLC), MalCodeAI promotes a proactive approach to secure coding practices. Additionally, its component-wise breakdown and reporting system ensures transparency and ease of interpretation, making security insights both accessible and developer-friendly.

\end{document}